**Identification by deuterium diffusion of a nitrogen-related deep donor preventing the p-type doping of ZnO**

N. Temahuki[1], F. Jomard[1], A. Lusson[1], I. Stenger[1], S. Hassani[1], J. Chevallier[1], J.M. Chauveau[1,2], C. Morhain[2], J. Barjon[1*]

[1] Université Paris-Saclay, UVSQ, CNRS, GEMaC, 78000, Versailles, France

[2] Université Cote d'Azur, CNRS, CRHEA, 06560, Valbonne Sophia Antipolis, France

Deuterium diffusion is investigated in nitrogen-doped homoepitaxial ZnO layers. The samples were grown under slightly Zn-rich growth conditions by plasma-assisted molecular beam epitaxy on m-plane ZnO substrates and have a nitrogen content [N] varied up to $5 \times 10^{18}$ at.cm$^{-3}$ as measured by secondary ion mass spectrometry (SIMS). All were exposed to a radio-frequency deuterium plasma during 1h at room temperature. Deuterium diffusion is observed in all epilayers while its penetration depth decreases as the nitrogen concentration increases. This is a strong evidence of a diffusion mechanism limited by the trapping of deuterium on a nitrogen-related trap. The SIMS profiles are analyzed using a two-trap model including a shallow trap, associated with a fast diffusion, and a deep trap, related to nitrogen. The capture radius of the nitrogen-related trap is determined to be 20 times smaller than the value expected for nitrogen-deuterium pairs formed by coulombic attraction between $D^+$ and nitrogen-related acceptors. The $(N_2)_O$ deep donor is proposed as the deep trapping site for deuterium and accounts well for the small capture radius and the observed photoluminescence quenching and recovery after deuteration of the ZnO:N epilayers. It is also found that this defect is by far the N-related defect with the highest concentration in the studied samples.

[*] corresponding author : julien.barjon@uvsq.fr

Zinc oxide (ZnO) is an attractive wide bandgap semiconductor with expected strong potentials in the fields of electronics, optoelectronics and spintronics. The lack of reliable p-type doping in ZnO has so far hindered its development in some of those fields. While n-type conductivity is easily obtained in ZnO, reliable p-type conductivity of ZnO has proven to be difficult to achieve in a reproducible way, if it has ever been achieved. The most promising acceptor species in ZnO appears to be nitrogen, which was shown to give rise to two acceptor levels in ZnO that are fully ionized at room temperature [1]. The shallowest level has a binding energy of ~165 +- 40 meV and was first identified from donor-acceptor recombination on the photoluminescence spectra of ZnO:N [2]. This level is also detected in capacitance-voltage measurements [1], along with a deeper acceptor, having a binding energy of 0.48 eV but large capture cross section. However, the concentration of these acceptor centers only accounts for less than 1% of the total chemical concentration [N] in the ZnO:N films, and the epilayers are usually found to be fully or almost fully compensated. It should be noted that most recent theoretical calculations predict that the substitutional $N_O$ level forms a deep acceptor center with an energy level 1.3 eV or deeper above the valence band, [3] while the deep acceptor model has been confirmed experimentally [4].

Deuterium/hydrogen diffusion has long proven to be a most helpful technique to help figuring out the mechanisms making doping inefficient and to explain compensation mechanisms [5] [6]. In this study, deuterium diffusion is investigated in the much-debated context of p-type doping of ZnO.

Nitrogen-doped ZnO epilayers were grown by molecular beam epitaxy (MBE) on m-plane $(10\bar{1}0)$ nonpolar ZnO substrates after the growth of a thin (Zn,Mg)O buffer layer, the role of which being to limit the diffusion of donor impurities from the substrates [7]. Experimental conditions similar to those of Ref. [8] were used, which resulted in non-intentionally doped (n.i.d.) ZnO epilayers exhibiting an extremely low contamination by residual impurities and a net residual donor concentrations as low as ~$10^{14}$ cm$^{-3}$ as measured by capacitance-voltage C(V) measurements, even without the use of a (Zn,Mg)O barrier layer. Three epilayers were grown and doped with different levels of nitrogen using a radiofrequency (r.f.) plasma cell fed with nitrogen gas. C(V) measurements of similar ZnO:N layers show that such epilayers remain n-type [1].

The samples were cut in two pieces by using a wire saw, in order to keep a reference of the as-grown state. To introduce deuterium in the samples, they were exposed during 60 minutes to a r.f. plasma of pure $D_2$, under 1 mbar and with 36W net excitation power at room temperature (~300 K). They were placed in remote position to avoid direct exposition to the energetic species of the plasma and ensuing defects. m-plane ZnO bare substrates were placed with the epilayers inside the plasma chamber during each deuteration. These reference samples were analyzed along with the deuterated ZnO:N layers to verify that the experimental conditions were rigorously the same from one deuteration to another.

Secondary ion mass spectrometry (SIMS) using a last-generation CAMECA IMS 7f equipment was performed to measure the concentration and depth distributions of deuterium and nitrogen atoms into the samples. $Cs^+$ primary ions were accelerated at 15 keV. The depth resolution was established to be less than 3.6 nm/decade (depth over which the signal drops by a factor of 10). This value was extracted from the slope of the magnesium signal at the interface between the (Zn,Mg)O buffer layer and the ZnO:N epilayer. and can be considered as an upper limit for the depth resolution. Nitrogen was measured by selecting mass 30 instead of mass 14 because it offers a better detection limit for nitrogen atoms in ZnO (about $1 \times 10^{17}$ $cm^{-3}$) due to the formation of $N^{14}O^{16}$ secondary ions. Nitrogen and deuterium absolute concentrations were quantified with a 10% uncertainty using implanted standards analyzed in the same conditions. The sample characteristics are presented in Table I.

Finally, photoluminescence (PL) spectra were recorded at low temperature (T=4K) under the excitation of a 325 nm He-Cd laser line with a low pump power set around 0.5 $W.cm^{-2}$. The as-grown and deuterated samples were mounted all together to ensure similar excitation and detection conditions, in order to compare directly their PL intensities.

Table I: Sample properties measured by SIMS: epilayer thickness, average nitrogen concentration [N], steepness of the deuterium diffusion front α and capture radius $R_c$ of nitrogen-related traps (details in the text).

| Sample | Description | Thickness (μm) | [N] (cm$^{-3}$) | 1/α (nm) | $R_c$ (Å) |
|--------|-------------|----------------|------------------|----------|-----------|
| A | n.i.d. ZnO epilayer | 1.1 | < 1x10$^{17}$ | - | - |
| B | nitrogen-doped ZnO epilayer | 1.1 | 1.0x10$^{18}$ | 15.3 | 3.4 |
| C | nitrogen-doped ZnO epilayer | 1.6 | 4.5x10$^{18}$ | 8.3 | 2.6 |

Figure 1 presents the deuterium and nitrogen SIMS depth profiles of the deuterated ZnO layers. Note that for the n.i.d. sample A, the nitrogen concentration is not reported since it lies below the detection limit (<1x10$^{17}$ cm$^{-3}$). In the nitrogen-doped samples, a sharp peak for mass 30 is observed at 1.1μm in sample B (respectively 1.6 μm in sample C). These peaks occur at the interface between the ZnO substrate and the (Zn,Mg)O buffer layer, thereby materializing the growth start. Note that these sharp peaks are not related to nitrogen, but rather to a silicon contamination of the substrate surface that led to an interference with secondary ions Si$^{28}$H$^2$ at mass 30. Otherwise, step-like doping profiles for nitrogen are obtained in both samples B and C. Looking in details at sample C data, after the beginning of the growth (i), the mass 30 signal lies at the nitrogen detection limit. When the nitrogen plasma source is lightened (ii), nitrogen is detected as a ~ 6x10$^{17}$ cm$^{-3}$ concentration step. The next increase to 4.5x10$^{18}$ cm$^{-3}$ occurs when the nitrogen source shutter is opened (iii). Altogether, these profiles illustrate how controlled the nitrogen doping can be in MBE grown ZnO epilayers.

Regarding the deuterium concentration profiles, it is necessary to remind that all samples were exposed to the same deuteration protocol. First, note that deuterium diffusion was effectively achieved at room temperature in all epilayers, which had not yet been evidenced before this work. Theoretical predictions [9] putting forward the high mobility of interstitial H in ZnO are thus confirmed by these present results.

The most important result provided by Fig. 1 is the following: the penetration depth of deuterium in the ZnO epilayers decreases as the nitrogen concentration increases (Fig. 1). In the n.i.d. sample A, deuterium diffuses over 1.1 µm *i.e.* close to the substrate, whereas for the nitrogen-doped samples B and C, the penetration depth is much smaller. D diffuses no further than ~0.5 µm in sample B doped with [N] = $1.0 \times 10^{18}$ at.cm$^{-3}$. This effect is further accentuated with the higher doping level of sample C, where D diffuses over only 0.2 µm from the surface. These SIMS profiles show unambiguously that the presence of nitrogen slows down deuterium diffusion in the ZnO crystal. This behavior is typical of a diffusion mechanism limited by the trapping of deuterium on a nitrogen-related defect. In other words, these results show that nitrogen-deuterium complexes are formed during the diffusion experiment at room temperature.

An additional result emerges from the detailed analysis of the deuterium profiles. For the n.i.d. sample A in Fig. 1(a), it is striking that the D diffusion stops precisely at the ZnO/(Zn,Mg)O interface, observable with lower ZnO signal. It evidences that the (Zn,Mg)O buffer layer provides an efficient diffusion barrier for deuterium, as already reported for aluminium in ZnO heteroepitaxy on sapphire [7]. Incidentally, (Zn,Mg)O layers might be regarded as a way to protect the particularly reactive and hydrophilic surface of ZnO.

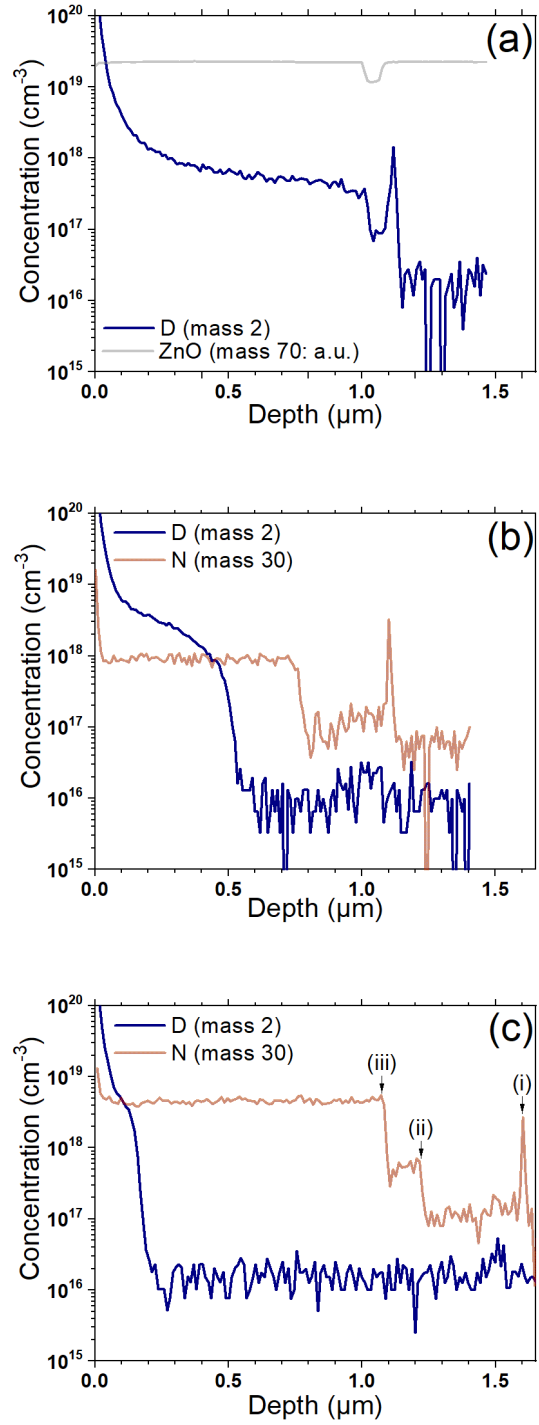

FIG.1: SIMS depth profiles after deuterium diffusion at room temperature in (a) the *n.i.d* ZnO sample A, where the ZnO matrix signal is also plotted in arbitrary units to evidence the (Zn,Mg)O barrier layer at ~1µm depth; (b) and (c) the nitrogen-doped samples B and C, respectively.

The deuterium profiles are further analyzed in the framework of a diffusion model with two types of trapping sites. This two-trap model was initially developed for describing hydrogen diffusion in amorphous [10] and polycrystalline [11] silicon. In Fig. 2, the deuterium profiles can be subdivided into 3 regions: (i) the first 50 nm below the surface present high concentrations of deuterium ($>10^{20}$/cm$^3$) due to its trapping on subsurface defects induced by the plasma process [12]. This region is not of great interest to analyze; (ii) at larger depths, we observe a fast diffusion regime for deuterium, well described by a complementary error function ($erfc$). This evidences the presence of shallow traps for deuterium atoms, shallow referring here to a weak binding between deuterium and the trap. When the deuterium concentration decreases until it reaches the nitrogen level, there is an abrupt change in the diffusion process; in region (iii), deuterium diffusion becomes trap-limited. These indications reveal the presence of two distinct types of trapping sites for deuterium: shallow traps, and deep traps involving nitrogen atoms.

When the deuterium concentration [D] exceeds the deep trap concentration, all of the deep-trapping sites are saturated. The excess deuterium diffuses very fast, following an intrinsic-like diffusion profile:

$$[D](x,t) = [D]_{0S}\, erfc\left(\frac{x}{\sqrt{4D_{eff}t}}\right)$$

Where x is the depth from the surface, t the duration of the exposition to the D plasma, $[D]_{0S}$ the deuterium concentration at the surface, and $D_{eff}$ the effective diffusion coefficient.

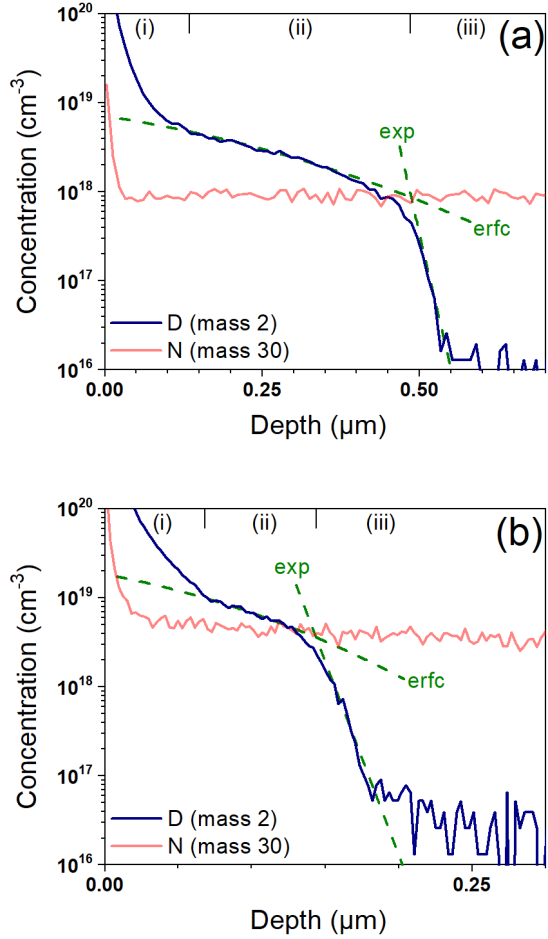

FIG.2: Complementary error function and exponential curve fitting (dashed lines) on D profiles for (a) sample B ($1\times10^{18}$ at.cm$^{-3}$) and (b) sample C ($5\times10^{18}$ at.cm$^{-3}$) nitrogen-doped ZnO epilayers. Note that the crossing point of the erfc and exp fits (green dashed lines) almost occurs at the value of the nitrogen concentration (red line).

In region (ii) where shallow traps are governing deuterium diffusion, the effective diffusion coefficient was found to be equal to $1.4\times10^{-13}$ cm$^2$.s$^{-1}$ in sample B. Note that in sample C the extent of region (ii) is too short to evaluate $D_{eff}$. Otherwise, the value obtained for $D_{eff}$ in sample B is similar to the one extrapolated from Ref. [13] at $1.4\times10^{-13}$ cm$^2$.s$^{-1}$ in the case of D diffusion in polycrystalline ZnO.

To further investigate the origin of the shallow traps, Fourier Transform Infra-Red (FTIR) absorption experiments were performed on n.i.d. ZnO single crystals which were deuterated during 4 hours at 200°C. With these conditions, higher quantities of deuterium were imbedded into the substrates. SIMS analysis shows erfc profiles (see supplementary material Fig. S1(a)) comparable to the fast diffusion regime seen in our ZnO:N samples. Interestingly, the infrared absorption line at 2466 cm$^{-1}$ was detected at 90K (see supplemental material Fig S1(b)). This value is close to the one found at 2470.3 cm$^{-1}$ at 8K in Ref. [14]. These authors identified the interstitial anti-bonding (AB) site on oxygen atoms to be a shallow trapping site for hydrogen in the ZnO lattice, which leads us to suggest that the AB interstitial location of deuterium is the shallow trapping site for deuterium in the fast diffusion regime observed in region (ii).

Region (iii), where [D] becomes lower than the deep trap concentration, marks the shift to a trap-limited diffusion regime. A negligible detrapping of deuterium can be assumed, consistently with the observed diffusion front which follows an *exp(-αx)* function. The capture radius R$_c$ of deuterium by the nitrogen-related traps can be assessed from the α value assessed with a linear regression in log-scale. 1/α is found to be equal to 15.3 nm and 8.3 nm for samples B and C respectively, much larger than the SIMS in-depth resolution (≤ 3.6 nm). The erfc and exp functions crosses almost exactly at the nitrogen concentration (Fig.2), which indicates that the deep trap concentration ≈ [N]. The capture radius of deuterium is then given by R$_c$ = α$^2$/(4π[N]) [15]. From this expression, capture radius values of 3.4 Å and 2.6 Å are deduced from samples B and C respectively, which is consistent given the experimental uncertainties.

Such a capture radius of ~3 Å appears surprisingly small. Indeed, in the case of a pure coulombic attraction between D$^+$ and N$^-$ leading to the formation of N-D pairs (D$^+$ + N$^-$ → N-D), the capture radius $R_c$ is obtained from the equivalency between Coulomb potential energy between charged particles of opposite signs (+q, -q) and thermal energy [15] : $R_c = q^2/4\pi\varepsilon kT$. In this formula, q is the elementary charge, ε the dielectric constant, k Boltzmann constant and T the temperature. This model is accurately verified for H captured by substitutional acceptors such as B in Si [16], Al in SiC [17] or boron in diamond [12]. However, with ε = 7x10$^{-11}$ F.m$^{-1}$ for ZnO [18], the equation yields $R_c$ = 70 Å at room temperature. This result is more than 20 times larger than the ~ 3 Å value assessed from the diffusion experiments. One could argue that a screening of the coulombic interaction in the presence of free carriers could reduce the capture radius. Following Ref. [19], more than $10^{21}$ free electrons/cm$^3$ would be needed to account for $R_c \sim 3$ Å. This is not consistent with the doping range of the investigated samples. We can therefore conclude that the mechanism of coulombic attraction between elementary charges (+q,-q) cannot explain the short capture radius of the nitrogen-related traps observed in this work.

While it seems reasonable to assume that deuterium is present under a D$^+$ positive state of charge in ZnO [20], the short capture radius indicates that nitrogen traps for deuterium *are not* present under the negative state of charge in the ZnO:N samples studied here. On the other hand, if the nitrogen-related defect involved in the trapping of deuterium is neutral, it becomes possible to account for the observed diffusion profiles: the binding of D$^+$ on a neutral center is governed by dipole interactions and then occurs at a much shorter distance range than the coulombic attraction. This scenario is consistent with the small values of the capture radius of deuterium by nitrogen-related traps measured in this work. As a conclusion, the diffusion profiles of deuterium in ZnO:N indicate that the nitrogen-related deep trap is under a neutral state of charge.

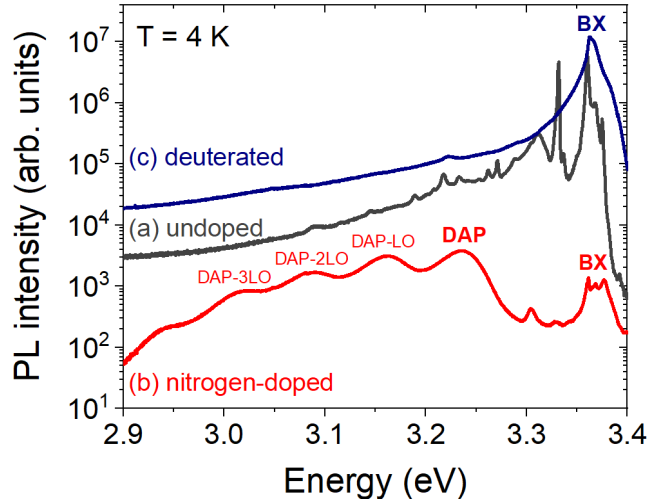

FIG.3: Compared PL intensities (no offset) at 4K of the (a) as-grown n.i.d epilayer A, (b) as-grown nitrogen-doped epilayer C and (c) deuterated nitrogen-doped epilayer C. The defect line at 3.334eV on the reference undoped sample might due to a non-optimal substrate preparation combined and exacerbated by the low pump power used here.

Figure 3 displays the effects of nitrogen doping and deuteration on the PL spectra. For the undoped sample A reported in Fig 3(a), the spectrum is dominated by bound exciton (BX) recombinations at ~3.36 eV as observed in high quality materials. When doped with nitrogen in Fig 3(b), donor-acceptor-pair (DAP) recombinations at ~ 3.23 eV dominate the PL spectrum, with longitudinal optical (LO) phonons replicas at lower energy, while the PL intensity of BX recombinations decreases by almost 4 orders of magnitude. The total PL intensity was further evaluated by integration from 2.9 to 3.4 eV which is significant since no deeper bands are observed at lower energy. The total PL intensity of the nitrogen-doped ZnO layer is smaller by a factor of 90 compared to that of the undoped sample. This indicates that nitrogen introduces in the ZnO epilayers a center which is strongly non-radiative. Finally, the most striking result concerns the deuteration of the nitrogen-doped epilayer in Fig.3(c). It is remarkable that the deuterated epilayer C recovers the PL intensity measured in the undoped epilayer A. Note that there is no evidence of the DAP disappearance after deuteration, since the band tail of near-band-edge luminescence is stronger. In fact, the deuterium diffusion mainly results in the passivation of the non-radiative nitrogen center responsible for the PL intensity quenching.

In this final part, let us discuss the nature of nitrogen-related centers which could lead to small capture radii for deuterium. As shown in the previous diffusion analysis, small capture radii dictate that the deuterium trapping center is neutral. In a n-type semiconductor, this situation can only occur if the trapping center is a deep donor. Therefore, none of the reported acceptor states could explain the observed deuterium diffusion profiles because in a n-type semiconductor, all acceptors are ionized by compensation. This holds for the 1.3 eV acceptor level of N substituting O [3], the 0.17 eV double acceptor level attributed to a molecular $N_2$ at zinc site $(N_2)_{Zn}$ [21, 22], the 0.48 eV acceptor level [1]. The first neighbor nitrogen pairs at oxygen sites $N_O$-$N_O$ has also been identified as deep acceptors, with binding energies slightly deeper than that of $N_O$ [23].

On the contrary the more-tightly bond nitrogen pairs $(N_2)_O$, *i.e* molecular nitrogen at the oxygen site, is not only expected to be a likely N-related defect due to its low formation energy under the zinc-rich conditions required to grow 2D ZnO epitaxial layers along this orientation, but also to be a deep double-donor [24, 25]. Such a deep donor is expected to be neutral in the presence of shallower deuterium donors with an ionization energy of 35 meV [26]. This scenario is fully consistent with our diffusion results. The passivation of the $(N_2)_O$ centers would involve the formation of $((N_2)_O, D_n)$ complexes, the atomic configuration of which remains to be studied theoretically. This scenario is also consistent with our optical results: the ZnO photoluminescence is quenched with nitrogen doping by a non-radiative center, presumably $(N_2)_O$. The PL intensity fully recovers after their passivation by deuterium, presumably as $((N_2)_O, D_n)$ complexes.

Finally, another key result is given by the change in diffusion mechanism occurring at almost identical deuterium and nitrogen concentrations on the SIMS profiles (Fig. 2). This reveals that the proposed $(N_2)_O$ defect is by far the nitrogen-related defect with the highest concentration in the studied samples and provides an answer to the long standing search for the dominant incorporation mechanism of nitrogen in ZnO:N grown under zinc-rich conditions.

In conclusion, deuterium diffusion carried out on m-plane homoepitaxial ZnO:N samples grown by MBE in slightly Zn-rich conditions and using a r.f. plasma nitrogen source, reveal that a large amount of the nitrogen incorporated in the epilayers is introduced as nitrogen-related deep donor defects. The $(N_2)_O$ defect with a rather low formation energy is a credible candidate for such a donor center. Its concentration has to be drastically reduced in favour of the nitrogen-related acceptors to achieve reliable p-type ZnO, which might be done by promoting O-rich rather than the usual Zn-rich growth conditions.

**Supplementary materials**

See supplementary materials for FTIR absorption experiments performed on undoped ZnO single crystals after deuteration.

**Data availability statement**

The data that support the findings of this study are available from the corresponding author upon reasonable request.